\documentclass[superscriptaddress,showpacs,aps,twocolumn,amsmath,amssymb,prl]{revtex4}

\usepackage{graphicx}
\usepackage{dcolumn}
\usepackage{bm}
\usepackage[latin1]{inputenc}          
\usepackage[english]{babel}

\begin{document}

\title{Ultrafast Acousto-Plasmonics in Gold Nanoparticles Superlattice}

\author{P. Ruello}
\email{pascal.ruello@univ-lemans.fr}
\affiliation{Institut des Mol\'ecules et Mat\'eriaux du Mans, UMR-CNRS 6283, Universit\'e du Maine, 72085 Le Mans, France}
\author{A. Ayouch}
\affiliation{Institut des Mol\'ecules et Mat\'eriaux du Mans, UMR-CNRS 6283, Universit\'e du Maine, 72085 Le Mans, France}
\author{G. Vaudel}
\affiliation{Institut des Mol\'ecules et Mat\'eriaux du Mans, UMR-CNRS 6283, Universit\'e du Maine, 72085 Le Mans, France}
\author{T. Pezeril}
\affiliation{Institut des Mol\'ecules et Mat\'eriaux du Mans, UMR-CNRS 6283, Universit\'e du Maine, 72085 Le Mans, France}
\author{N. Delorme}
\affiliation{Institut des Mol\'ecules et Mat\'eriaux du Mans, UMR-CNRS 6283, Universit\'e du Maine, 72085 Le Mans, France}
\author{S. Sato}
\affiliation{School of Science, University of Hyogo, 3-2-1 Koto, Kamigori-cho, Ako-gun, Hyogo 678-
1297, Japan}
\author{K. Kimura}
\affiliation{School of Science, University of Hyogo, 3-2-1 Koto, Kamigori-cho, Ako-gun, Hyogo 678-
1297, Japan}
\affiliation{Research Center for Micro-Nano Technology, Hosei University, Japan}
\author{V. Gusev}
\affiliation{Laboratoire d'Acoustique, UMR CNRS 6613, Universit\'e du Maine, 72085 Le Mans, France}




\begin{abstract}
We report the investigation of the generation and detection of GHz coherent acoustic phonons in plasmonic gold nanoparticles superlattices (NPS). The experiments have been performed from an optical femtosecond pump-probe scheme  across the optical plasmon resonance of the superlattice. Our experiments allow to estimate the collective elastic response (sound velocity) of the NPS as well as an estimate of the nano-contact elastic stiffness. It appears that the light-induced coherent acoustic phonon pulse has a typical in-depth spatial extension of about  45 nm which is roughly 4 times the optical skin depth in gold. The modeling of the transient optical reflectivity indicates  that the mechanism of phonon generation is achieved through ultrafast heating of the NPS assisted by light excitation of the volume plasmon. These results demonstrate how it is possible to map the photon-electron-phonon interaction in subwavelength nanostructures. 

\end{abstract}

\maketitle

Plasmon assisted subwavelength light transmission is a key physical mechanism for modern nano-optics and nano-plasmonics \cite{ebessen}. The propagation of plasmon-polariton waves in nano particles nanostructures (chains, arrays) is at the core of intense fundamental investigations and has led to numerous reports \cite{brong,maier,maier2,solis,cemes}. For plasmonic applications, it is obviously crucial to understand and control the damping of these plasmon-polariton waves, i.e. the collective propagation distance. This propagation distance  as well as a more general description of the dispersion curves for both longitudinal and transverse light electric field have already been obtained for supported 1D chains  \cite{brong,maier,maier2,solis}.  The propagation distance is currently limited by the intrinsic electron-phonon, electron-defect interaction within each particle as well as radiation loss. In some particular 2D optical spectroscopy imaging, it has been shown recently that the propagation/distribution of the plasmon-polariton could even be mapped for 1D nanoparticles chains \cite{solis,cemes}. The visualization of these plasmons at the interface air/nanostructure with near-field imaging \cite{maier2} or with an electron beam imaging \cite{bosman} were also reported. This state-of-art 2D imaging is however limited to 1D or 2D systems, i.e. to surface plasmonics and no local measurement of volume plasmon-polariton propagation has been reported for 3D nanoparticles arrays so far. Most of the plasmonic responses in nanoparticles superlattices are obtained indeed from far field optical absorbance measurements \cite{kimuraJAP,tao}. Because of the difficulty to probe the innner part of a plasmonic superlattices, no quantitative depth profiling description of the plasmon-polariton propagation have been reported to date. However, it is known that ultrafast optical techniques can provide a picture in time and space of the light-matter interaction \cite{tom1,gusev1,ruelloul}. In particular, in case of femtosecond light pulses, the light-matter coupling can lead to the emission of coherent acoustic phonons that occurs  all over the spatial extension where the light energy is converted into mechanical energy. Consequently, the characteristic spatial profile of the emitted coherent acoustic phonons contains information on the light-matter energy conversion process. In bulk solids, the  analysis of coherent acoustic phonons packet has permitted to understand the evolution in time and in space of the different electron-acoustic phonons interactions like the thermoelastic process (ultrafast heating of the lattice) \cite{tas,wri4,lejmanB}, the deformation potential process (electronic pressure) \cite{chi,wri3,young} and even, reported recently, the spatial electron-hole separation in semiconductors (i.e. the photoinduced Dember electric field)  \cite{vaudel}. 
\begin{figure*}[t]
\centerline{\includegraphics[width=18cm]{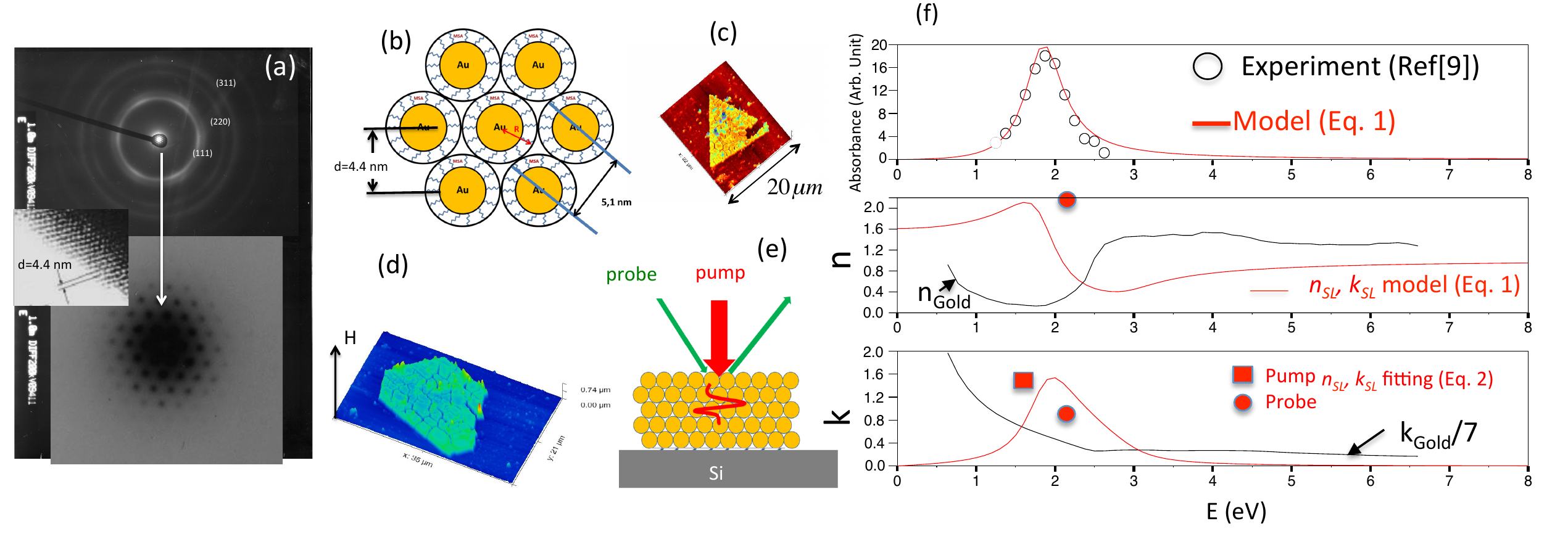}}
\caption{\label{fig1} (color online)  (a) Transmission electron diffraction (TED) pattern showing the Debye Scherrer rings (top figure) coming from the crystallized cubic gold nanoparticles. In the middle figure, the transmission electron microscopy image (TEM) shows the mesoscopic hexagonal arrangement (hcp) with the two neighbor plane distance of 4.4 nm. The hcp arrangement is also well evidenced with small angle electron diffraction revealing the 6-fold axis (bottom figure). (b) Sketch of the gold nanoparticles superlattices connected via MSA molecules. (c) Optical microscopy view of a superlattice  deposited onto a silicon substrate. (d) AFM profile of the hexagonal shape superlattice (height H=180 nm). (e) Sketch of the pump and probe experiments performed on these nanoparticles superlattice where coherent acoustic phonons are generated and detected in the front-front configuration. (f) Top figure : plasmon resonance of the gold superlattice from\cite{kimuraJAP} fitted with Eq. (\ref{eq1}). Middle and bottom figures : refractive index (real and imaginary parts) of the gold superlattice compared to those of bulk gold \cite{johnson}. The red symbols (square and circles) are the refractive index values  estimated with the photoacoustic response (see Eq. (\ref{drr}) and details in the text).}
\end{figure*}

In this Letter, by applying optical pump-probe scheme depicted in Fig. \ref{fig1}, we report the observation of GHz coherent acoustic phonons generation in NPS. The laser excited coherent acoustic pulse has a spatial extension  in the range of 45 nm which is much larger than either the gold nanoparticle diameter (3.7 nm) or the light penetration depth in gold.  Because laser-excited hot electrons are confined in the gold nanoparticles, these observations indicate that the acoustic phonons are very likely induced by ultrafast heating of the superlattice assisted by a collective plasmon-polariton propagation. While individual vibrations of nanoparticles assemblies have been reported earlier \cite{bigot,mante} or narrowband Brillouin mode in semitransparent cobalt superlattice \cite{dario}, our results  demonstrate that it is possible to generate propagating broadband coherent acoustic phonons in plasmonic NPS. 

The samples have been grown by soft chemical routes  (for ample details see the review \cite{kimura1}). These nanoparticles, encapsulated with 0.7 nm long molecular chains (MSA Mercaptosucinic), are packed on a silicon substrate according to an hexagonal structure (hcp-packing) with the $c$ axis perpendicular to the substrate (the 6-fold symmetries are clearly visible in Fig. \ref{fig1}(a)-(c)-(d) from mesoscopic to microscopic scales). The measured NPS spacing of $d$ = 5.1 nm, gives the particle diameter of 3.7 nm. Two different superlattices have been studied with typical lateral size of tens of micrometers and a thickness of 52 nm (triangular shape, Fig. \ref{fig1}(c)) and 180 nm (hexagonal shape, Fig. \ref{fig1}(d)). These ordered assemblies exhibit plasmonic properties (Fig. \ref{fig1}(f)) with a collective response modeled with the following simple phenomenological dielectric function \cite{cox,kimurachem}: 
\begin{eqnarray}
\label{eq1}
\epsilon_{SL}(\omega) = 1+\frac{\omega_{p}^{2}}{ \omega_{s}^{2} -\omega_{p}^{2}/3-\omega^{2}-i\Gamma\omega} 
 \end{eqnarray}
The real ($n_{SL}$) and imaginary ($k_{SL}$) parts  of the refractive index, as well as the plasmon dielectric losses ($\propto 2nk\omega$ \cite{ziman}) have been deduced from this dielectric function ($\tilde{n}_{SL} = n_{SL}+ik_{SL}=\sqrt{\epsilon_{SL}}$) and is shown in Fig. \ref{fig1}(f) where $\omega_S$=2.3 eV, $\omega_P$=2.2 eV and $\hbar\Gamma$=0.59 eV, are the surface plasmon resonance energy of the individual nanoparticle, the effective volume plasmon energy  and its damping energy respectively. One can notice that both the real and imaginary parts are clearly different from those of bulk gold crystal \cite{johnson} (see middle and bottom panels in Fig. \ref{fig1}(f)). By comparing this model and more recent numerical calculations \cite{dionne}, we can notice that Eq. (\ref{eq1})  gives a reasonable approximation for very small nanoparticles (for instance 3.7 nm) while it does  not for larger nanoparticles diameter ($>$ 20 nm).

The pump probe technique used here is based on a 80 MHz repetition rate Ti:sapphire femtosecond laser. The beam is split with a polarizing beam splitter into a pump and a probe beams. The probe beam is introduced in a synchronously pumped OPO that allows to tune the wavelength and permit to perform two-color pump-probe experiments (pump and probe are linearly and circularly polarized respectively). The transient optical reflectivity signals are obtained thanks to a mechanical delay stage (delay line) which enables a controlled arrival time of the probe pulse regarding to the arrival of the pump pulse. The experiments were conducted with incident pump and probe beams perpendicular to the surface as shown in Fig. \ref{fig1}(e).  

\begin{figure}[t!]
\centerline{\includegraphics[width=8.5cm]{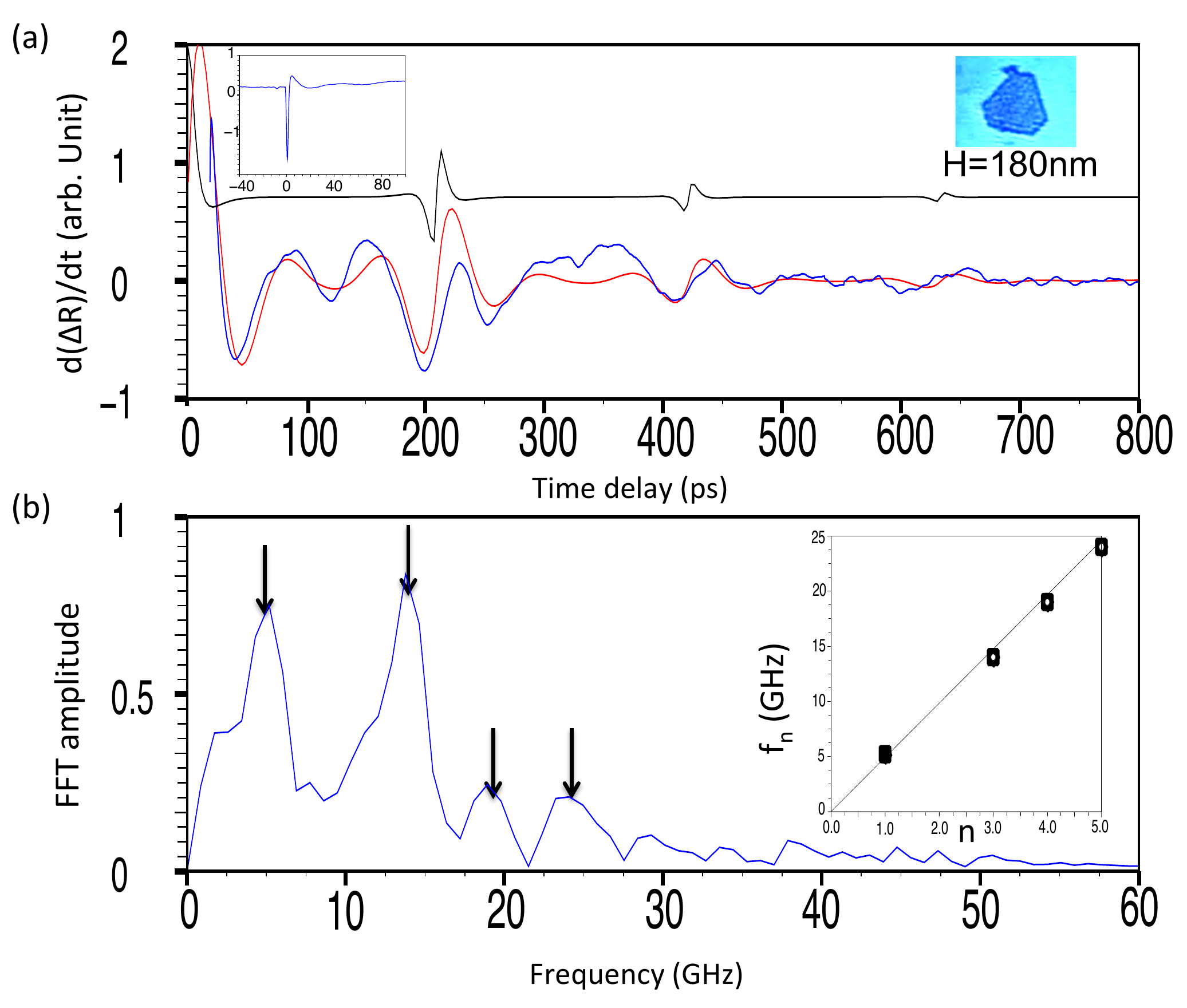}}
\caption{\label{fig2} (color online) (a) Time-derivative of transient optical reflectivity revealing the coherent acoustic phonon signal (blue curve) with numerical adjustment (red curve) containing information on the acousto-plasmonic processes of the generation and detection of the acoustic phonons (see Eq. \ref{drr}) (inset shows the fast electronic response of the superlattice). As a comparison a simulation of the coherent acoustic phonons signal is given when the pump light energy is converted only within the gold crystal skin depth (i.e. 11 nm at 1.59 eV without hot electrons diffusion) and probed with standard optical properties of gold crystal (black curve). (b) Corresponding coherent acoustic phonons spectrum obtained by a Fast Fourier Transform (FFT) of the experimental time derivative transient signal shown in (a). Four harmonics of the mechanical resonance of the superlattice appear (inset). }
\end{figure}

\begin{figure}[t!]
\centerline{\includegraphics[width=9cm]{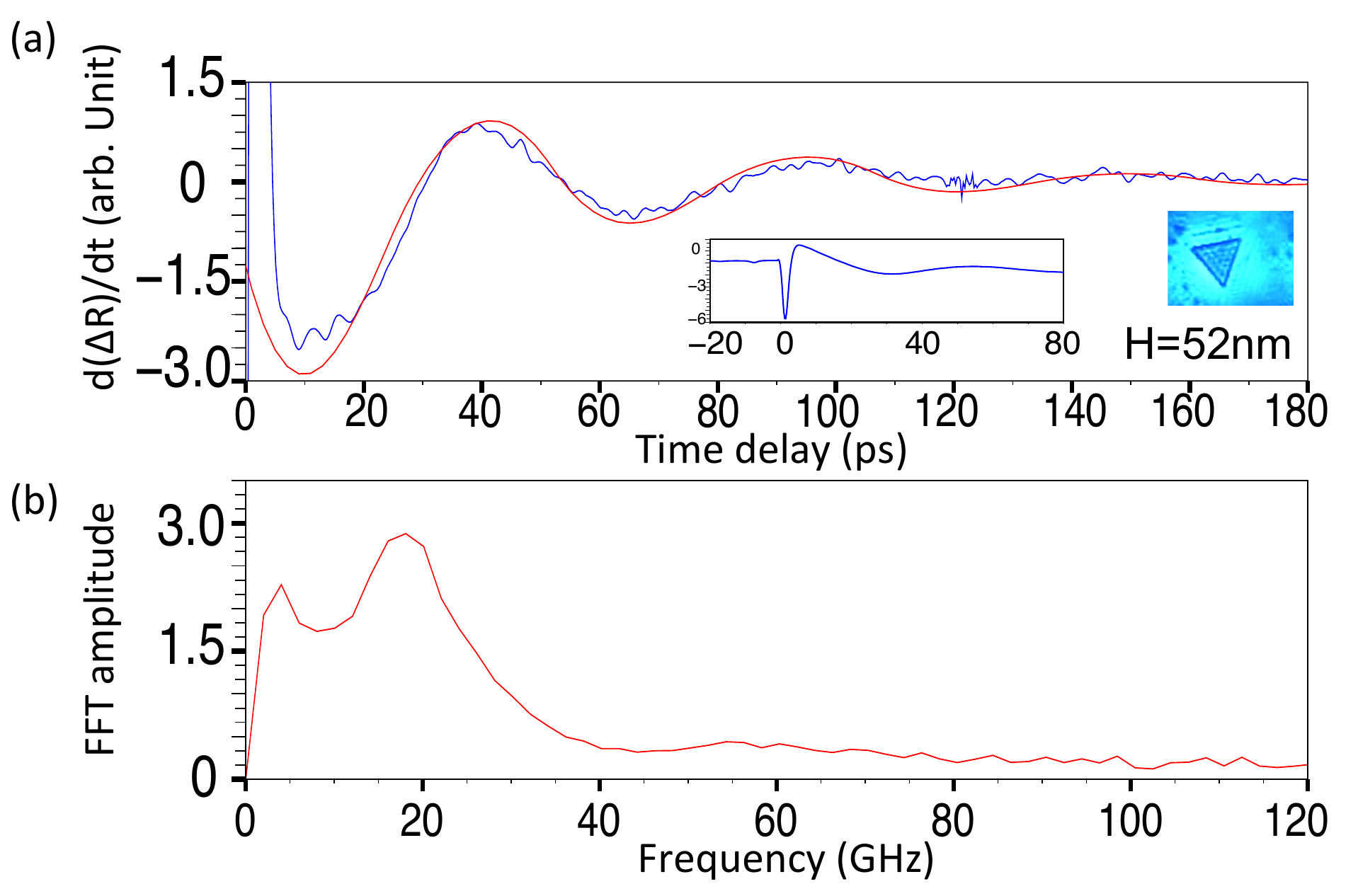}}
\caption{\label{fig3} (color online) (a) Time-derivative of transient optical reflectivity revealing the coherent acoustic phonon signal (blue curve) with numerical adjustment with a damped sinus (red curve) (inset shows the fast electronic response of the superlattice). (b) Corresponding coherent acoustic phonons spectrum obtained by a Fast Fourier Transform (FFT) of the experimental time derivative transient signal shown in (a).}
\end{figure}

The transient optical reflectivity signals obtained for superlattice are shown  in Figs. \ref{fig2} and \ref{fig3}. 
Time resolved reflectivity signals have been recorded for variable probe wavelengths (2.17-2.25 eV)  but no significant variation have been observed in this wavelength range, probably because of the broad collective plasmon resonance (Fig. \ref{fig1}(f)). 
For the thicker  NPS of 180 nm, the transient optical reflectivity signal  exhibits oscillatory components up to nearly 1 ns (pump 1.59 eV, probe 2.19 eV). In order to better reveal the oscillatory parts of the signal we  processed the numerical time derivative shown in Fig. \ref{fig2}(a). At first glance, the observation of the signal in the time domain reveals some periodic bursts  repeated about every 200 ps (Fig. \ref{fig2}(a)). We attribute this to an acoustic echo laser generated  at the NPS free surface  and travelling back and forth  across the NPS and periodically detected at the front side upon reflection at  the Si substrate interface. This interpretation  is confirmed  from the analysis of the FFT signals (Fig. \ref{fig2}(b)) where the detected frequencies follow the expected eigenmodes frequency sequence $f_{n}=nV_{LA}/2H$, with n=1, 2, 3... and $V_{LA}$ is the longitudinal sound velocity and $H$ the NPS thickness (see inset of Fig. \ref{fig2}(b)). From  the mean least square analysis of the slope, we extract a sound velocity of $V_{LA}$=1770 m.s$^{-1}$ which  is  about half of that measured in bare gold crystals. The decrease of the sound velocity  is an indication of the soft nanocontacts between nanoparticles. The time derivative signal shown in Fig. \ref{fig3}(a) corresponding to the thinnest NPS of 52 nm thickness  resembles a damped sinus function of about 18-19 GHz fitted frequency (Fig. \ref{fig2}(b)). As already observed in different nanometric thin films \cite{mechri,ayouch,pezleon}, since  the  NPS is much thinner (52 nm) than the previous one (180 nm), the photoexcitation leads to the mechanical resonance of the entire superlattice layer  at the first predominant harmonic frequency $f_1$  which leads to the estimate of V$_{LA}$ $\approx$ 1870-1970 m.s$^{-1}$, in close agreement with the previous estimate obtained for the thicker superlattice. The estimate of the sound velocity provides a direct evaluation of the molecular contact elastic stiffness. Following the hcp lattice dynamic equation along the z axis, we can connect the sound velocity to the nanoparticle contact effective stiffness $K$ with V$_{LA}$=$(4/\sqrt{3})d\sqrt{K/m}$ \cite{merkel} where $d$=5.1 nm is the interparticle distance.  With a particle diameter of 3.7 nm the mass becomes $m$=5.1.10$^{-22}$ kg, leading to $K$ $\approx$ 11 N.m$^{-1}$ consistently with covalent bonds and larger than Van der Waals connection probed in soft assemblies of nanoparticles \cite{ayouch,mante}. 

Beside the evaluation of collective sound velocity, the analysis of the coherent acoustic phonons spectrum provides insights on the ultrafast generation processes. The transient optical signal reported for the thicker superlattice is clearly different from those obtained earlier for films made of noble metals like gold, silver \cite{wri4} or copper \cite{lejmanB}. 
The physical characteristics of the photogeneration and photodetection processes can be obtained thanks to a proper simulation of the transient reflectivity signal ($\Delta$R/R). We performed this simulation for the thicker superlattice. We employed the standard transient reflectivity model where in our case $\Delta$R/2R=Real($\delta$r/r) with \cite{tom1,gusacta}:

\begin{eqnarray}
\label{drr}
\delta r/r&\simeq&\rho + i\delta\phi\simeq -2ik_{0}\delta z \\
&+& \frac{4ik_{0}\tilde{n}_{SL}}{(1-\tilde{n}_{SL}^{2})}\frac{d\tilde{n}_{SL}}{d\eta}\int_{0}^{\infty}\eta(z,t)exp(2ik_{0}\tilde{n}_{SL}z)dz \nonumber.
\end{eqnarray}

\begin{figure}[t!]
\centerline{\includegraphics[width=9cm]{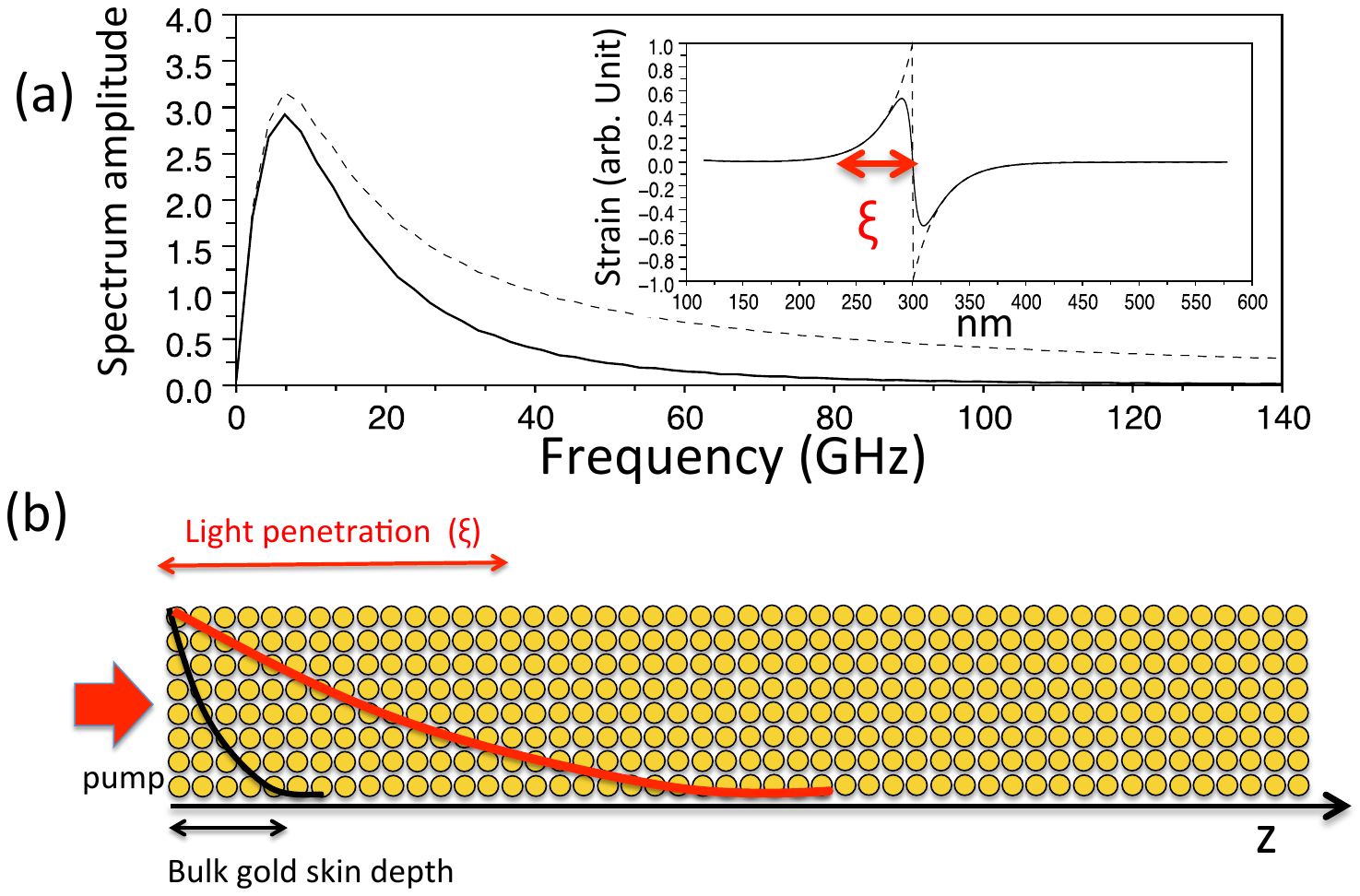}}
\caption{\label{fig4} (color online) (a) The full (dotted) line curve represents the photoinduced coherent acoustic phonons spectrum with (without) phonon attenuation used to adjust the transient photoacoustic response shown in Fig. \ref{fig3}(b).  The inset shows the emitted coherent acoustic phonon strain (ultrafast heating of nanoparticles) with a spatial extension of 42 nm used for the simulation. This spatial extension is sketched by the red curve in (b) where it is compared to the penetration depth of light into bulk gold.}
\end{figure}

where $k_{0}=2\pi/\lambda$ is the probe wave vector in air, $\eta(z,t)$ is the coherent acoustic phonons strain field propagating perpendicularly to the surface of the sample and $d\tilde{n}_{SL}/d\eta$ is the photoelastic coefficient. $z$=0 corresponds to the air/superlattice interface. In Eq. \ref{drr}, the first term corresponds to the contribution of the surface displacement ($\delta z$) of the superlattice and the second one to the photoelastic contribution (modification of the refractive index induced by the strain field of the coherent acoustic phonons). The strain field has been modeled by applying the standard bipolar shape \cite{tom1} with  $\eta(z,t)=-sgn(z-V_{S}t)exp(-\mid{ z-V_{S}t}\mid / \xi)$. The simulation of the photoinduced strain $\eta$(z,t) is shown as a dotted curve in inset of Fig. \ref{fig4}(a) where $\xi$ is the  characteristic depth over which the photo-induced strain occurs.  $\xi$ is related to the apparent imaginary part of the refractive index $k_{SL}^{pump}$ at the pump wavelength with  $\xi$=$\lambda$/4$\pi k_{SL}^{pump}$. In order to take into account phonon attenuation (anharmonic damping, defect scattering), we have used a cut-off frequency at around 30 GHz (full line curves in Fig. \ref{fig4}(a) and inset of Fig. \ref{fig4}(a)) in accordance with the experimental FFT (Fig. \ref{fig2}(b)). The calculation takes into account the reflection of the acoustic phonon wavepacket  at the silicon  interface to reproduce the three visible acoustic echoes. The best adjusted parameters we  have obtained  concerning the probe optical properties (@ 2.19 eV) are $\tilde{n}_{SL}^{probe}$=2.2+0.9i and for the pump (@ 1.59 eV) $k_{SL}^{pump}$=1.5. These  optical coefficients are  plotted in Fig. \ref{fig1}(f) and the fitted curve of the time derivative $d(\Delta R/R)/dt$ compared to the experimental  data is shown in Fig. \ref{fig2}(a). The adjustment of the shape of the photoacoustic signal depends also on the photoelastic coefficient $d\tilde{n}_{SL}/d\eta$ \cite{tom1} which  are unknown for such artificial nanoparticles superlattices. However we have tested different  scenarios and our adjustment shows that the real part of the photoelastic coefficient is dominant over  its imaginary part. Considering the well known relationship $d\tilde{n}_{SL}/d\eta$ $\propto$  $d\tilde{n}_{SL}/dE$ \cite{tom1}, $E$ being the probe energy, and from the energy dependence of optical refractive of the NPS refractive index shown in Fig. \ref{fig1}(f), it appears that the derivative of the real part ($n_{SL}$) (in particular @ 2.19 eV) is much larger than that of the imaginary part ($k_{SL}$). Consequently and in accordance with our simulations, it indicates that the real part of the photoelastic coefficient may dominate with $d\tilde{n}_{SL}/d\eta$ $\sim$ $dn_{SL}/d\eta$.

The results of  our simulations confirm that the optical properties of the  NPS are deeply different than those of bulk gold crystal. As a comparison, we have performed as well a calculation where we simulate the transient optical signal considering that the photo-induced strain occurs over a depth of 11 nm beneath the surface, as in bare gold metal as sketched in Fig. \ref{fig4}(b) ( hot electrons diffusion is neglected). The simulation is shown in Fig. \ref{fig2}(b) (black curve) obtained with an optical refractive index for gold (@2.19 eV) $\tilde{n}_{Au}$=0.25 + 3$i$ and (@1.59 eV) $\tilde{n}_{Au}$=0.3 + 6$i$. The discrepancy between this simulation and our experimental result clearly confirms that the pump light can deeply penetrate this subwavelength plasmonic nanostructure over a distance of about  $\xi\sim$ 42 nm, much larger than the optical penetration depth of bare gold. This distance over which the conversion of light energy into mechanical energy occurs is not compatible with transport of hot electrons due the existence of insulating contacts in between the nanoparticles as already discussed in \cite{kimurachem}. As a consequence, the ultrafast generation of coherent acoustic phonons in our NPS can only be interpreted from an ultrafast heating of the NPS assisted by transport of the electromagnetic energy through plasmon-polariton propagation perpendicular to the surface. 

As a summary, we have demonstrated the generation and detection of coherent acoustic phonons in gold nanoparticles superlattices. The analysis of the GHz sound propagation has permitted to estimate the nanocontact stiffness. Our results  demonstrate that the emission of coherent acoustic phonons occurs in the  NPS over a distance of about 45 nm. The  spreading of the conversion of light energy into mechanical energy requires transportation of the light energy in the subwavelength regime. We attribute this effect to an ultrafast heating assisted by plasmon-polariton propagation perpendicular to the NPS surface. The model of the generation and detection of coherent acoustic phonons is consistent with this mechanism and the optical parameters extracted from our fitting are in agreement with  conventional  optical characterizations. These results open new perspectives for the manipulation of plasmons by coherent phonons or  viceversa as recently investigated both theoretically and experimentally \cite{daniel,vasily}. 

We thank Dr. Yao for helpful discussion on the structural data of the superlattice. We thank Dr. V. Temnov and Dr. D. Lanzillotti-Kimura for fruitful discussions on plasmonics. This work was supported by the French Ministry of Education and Research, the CNRS and R\'egion des Pays de la Loire (CPER Femtosecond Spectroscopy equipment program).

\newpage

\end{document}